# A Compact Dual Band Dielectric Resonator Antenna For Wireless Applications

H. Raggad[1], M. Latrach[1], A. Gharsallah[2] and T. Razban[3]

[1]Radio-Frequency and Microwave Research Group, ESEO10 Boulevard Jeanneteau – CS 90717 - 49107 Angers Cedex 2, France
hedi.raggad@eseo.fr
[2]Unit of Research in High Frequency Electronic Circuits and Systems Faculty of Science of Tunis, El Manar, Tunisia
[3]IETR, Ecole polytechnique de l'Université de Nantes, BP 50609, 44306 Nantes, France

*ABSTRACT*

*This paper presents the design of a dual band rectangular Dielectric Resonator Antenna (DRA) coupled to narrow slot aperture that is fed by microstrip line. The fundamental $TE_{111}$ mode and higher-order $TE_{113}$ mode are excited with their resonant frequencies respectively. These frequencies can be controlled by changing the DRA dimensions. A dielectric resonator with high permittivity is used to miniaturize the global structure. The proposed antenna is designed to have dual band operation suitable for both DCS (1710 - 1880 MHz) and WLAN (2400 - 2484 MHz) applications. The return loss, radiation pattern and gain of the proposed antenna are evaluated. Reasonable agreement between simulation and experimental results is obtained.*

## 1. INTRODUCTION

With the increasing demand for high performance communication networks and the proliferation of mobile devices, sign cant advances in antenna design are essential in order to meet tomorrows requirements. In recent years, the rising development of the wireless communication industry have pushed antenna technology for increased performance. while being limited to an ever decreasing footprint. Such design constraints have forced antenna designers to multi band antennas. In the last decade, the attempts have been made to design dual band antennas using the dielectric resonator (DR). Dielectric resonator antenna (DRA) [1] [2] has been of interest due to their low loss, high permittivity, light weight and ease of excitation. In addition, wide bandwidth, and high radiation efficiency are inherent advantages of DRAs. Deferent shapes of DRAs such as cylindrical, hemispherical, elliptical, pyramidal, rectangular, and triangular have been presented in the literature. The rectangular-shaped DRAs over practical advantages over cylindrical and hemispherical ones in that they are easier to fabricate and have more design flexibility [3].

Various dual mode DRAs have been studied in recent years. For example, two DRA elements resonating at two deferent frequencies were combined to give a dual band antenna [4]. The resonant mode of the feeding slot was also utilized to widen the bandwidth [5]. Recently, a higher-order mode of a DRA was used to increase the bandwidth [6] or to design a dual band

DRA [7]. This higher-order-mode design approach has two major advantages. First, it requires no additional DRA and second, it facilitates the matching of the dual band antenna.

In this paper, a miniaturized dual band rectangular DRA is presented. In the proposed DRA both the fundamental $TE_{111}$ and the higher-order $TE_{113}$ propagation modes are excited. Moreover, in order to cover the DCS and the WLAN bands, the resonant frequencies of the $TE_{111}$ and $TE_{113}$ modes are adjusted by changing the dimensions of the DRA. The reflection coefficient, radiation patterns, and antenna gain of the dual band DRA were simulated with CST and HFSS simulators. A good agreement between the calculated resonant frequencies and the measurement results has been obtained.

## 2. ANTENNA CONFIGURATION

Figure 1 and 2 show the geometric configuration of a slot-coupled DRA fed by a microstrip line. The FR4 substrate with a thickness t = 0.8 mm and a permittivity of 4:4 is used. The ground-plane with an etched slot and an area of $70 \times 70$ mm$^2$ is located on the top surface of the substrate. On the bottom a microstrip line of 2.6 mm width is placed. The end of the microstrip line is extended beyond the slot center to maximize the coupling and to serve as a matching circuit. The dielectric resonator with permittivity of 30 is optimized and the determined dimensions are: a = 22 mm, b = 12.5 mm, and h = 17 mm, $L_s$ = 11 mm, $W_s$ = 2.6 mm for the slot, and L = 5 mm for the extension of the microstrip line.

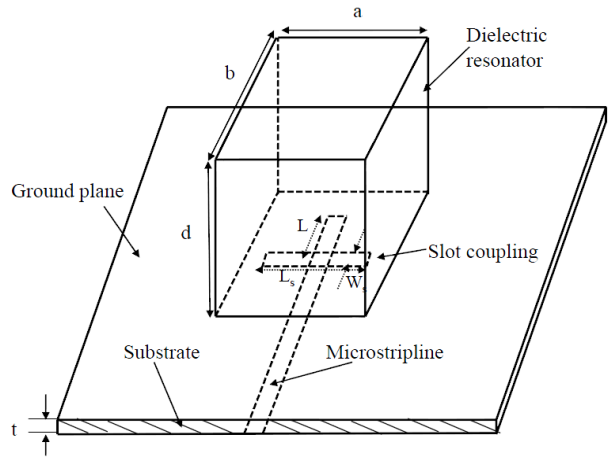

Figure 1. A rectangular Dielectric resonator antenna geometry.

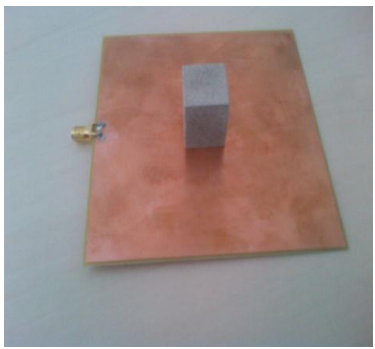
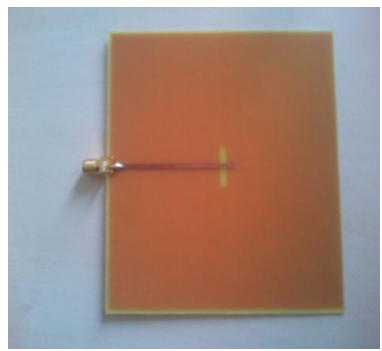

(a)                                        (b)

Figure 2. Fabricated Dielectric Resonator Antenna

## 3. PARAMETRIC STUDY

According to figure 1, there are number of parameters that influence on the conical bowtie DRA antenna performances. A parametric study is carried out to determine the effect of dielectric resonator permittivity, slot and microstrip length on the DRA characteristics.

### 3.1. DR Permittivity

It is well known that the resonant frequencies of high dielectric permittivity loading antenna are lower than the low dielectric loading one. Figure 3 shows the simulated reflection coefficient with different DR permittivity's (28, 30 and 32). It can be observed that the resonant frequencies decrease as the permittivity increases. As a result, a permittivity of 30 is used to design the proposed DRA, leading to an input return loss (S11) of -18dBm and -22dBm for $f_1$ and $f_2$ respectively.

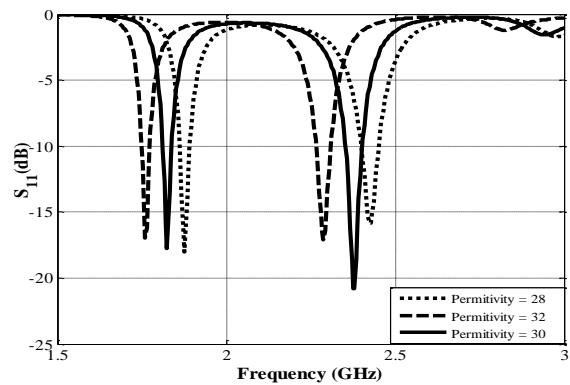

Figure 4. Return Loss vs. Frequency for different permittivity ($\varepsilon_r$)

### 2.3. Slot Length

Figure 4 shows the simulated return loss with different slot length ($L_{slot}$). When varying $L_{slot}$ from 12 mm to 15 mm an S11 degradation of 9 dB is observed at the second resonant frequency (2.4 GHz).

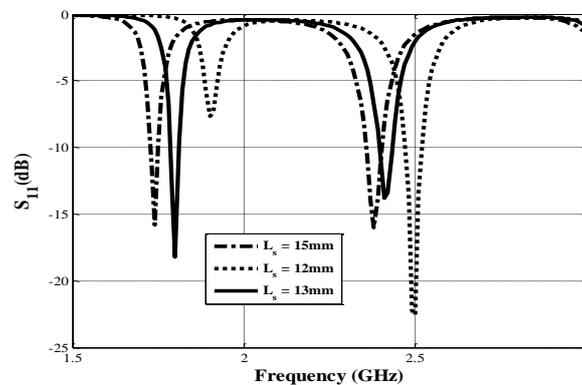

Figure 5. Return Loss vs. Frequency for different stub length ($L_s$)

## 2.4. Microstrip Length

Figure 5 shows the reflection coefficient as a function of frequency for different stub lengths. It can be seen that tuning the stub length only slightly affects the input matching of the resonant modes. The optimized length chosen for our design is L = 5 mm.

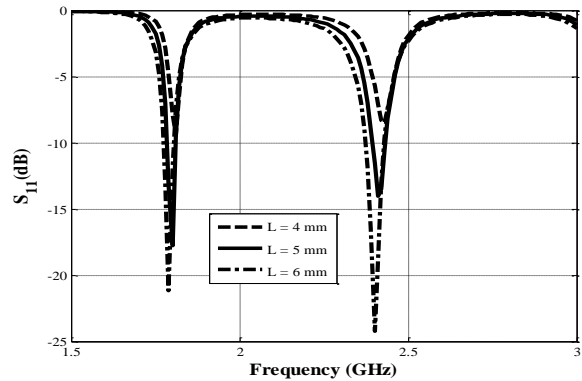

Figure. 5. Return Loss vs. Frequency for different stub length (L)

## 3. SIMULATION AND MEASURMENT RESULTS

An antenna prototype was fabricated using FR4 substrate as shown in Figure1 . Figure 6 illustrate the measured and simulated return loss of the dualband antenna, which is plotted together with the simulated results for comparison purpose and reasonable agreement between them is observed. The measured resonant frequencies of the lower and upper bands are 1.80 GHz and 2.43 GHz respectively, which agree very well with the HFSS simulated frequencies of $f_1$ = 1.82 GHz (0.025 % error) and $f_2$ = 2.4 GHz (1.25 % error) and CST simulated frequencies of $f_1$ = 1.85 GHz (2.2 % error) and $f_2$ = 2.38 GHz (2.08 % error). The measured bandwidths of the lower and upper bands are 8.3 % (1.72 - 1.87 GHz) and 3.7 % (2.39 - 2.48 GHz), respectively, covering the entire DCS and WLAN bands. Figure 7 shows the maximum measured gain versus frequency for the proposed antenna. The curve exhibits two peak gain of 4.7dBi and 4.2dBi at 1.8 GHz and 2.4 GHz respectively. Figure 8. shows the measured radiation patterns of the proposed DRA at the two $TE_{111}$ and $TE_{113}$ mode frequencies. The figure, reveals that the two resonant modes exhibit broadside radiation patterns and are very similar to each other, which is desirable. For each resonant mode, the co-polarization field are stronger than the cross-polarized counterparts by more than 20 dB in the broadside direction (θ° = 0). A good asymmetry radiation pattern in the two plans E and H is observed. However, an asymmetry in the H plane $TE_{111}$ mode pattern is remarked at 30°. This shows that the effect of the matching slot on radiation field is not significant.

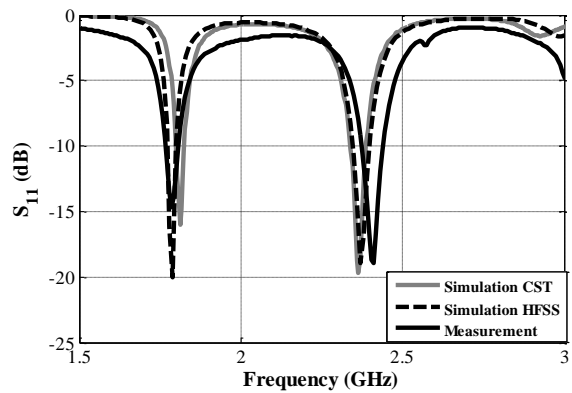

Figure 6. Measured and simulated Return Loss

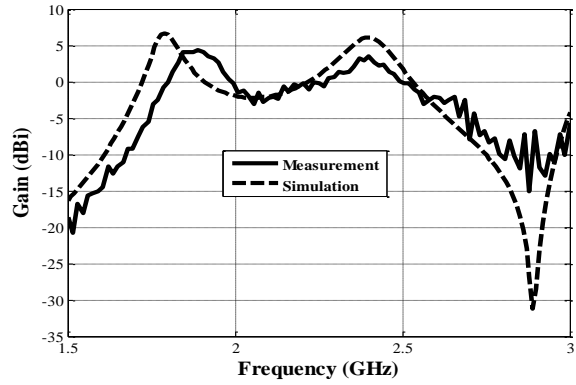

Figure 7. Measurement gain versus

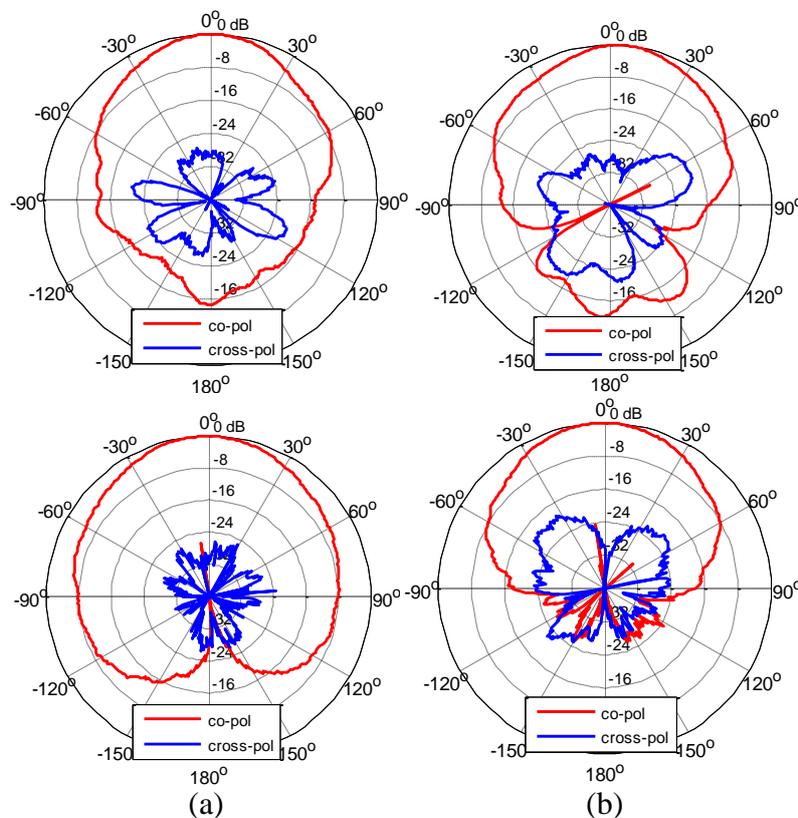

(a)            (b)

Figure.9. Measured radiation patterns of the $TE_{111}$ ( 1.8 GHz) and $TE_{113}$ (2.4 GHz) modes.

(a) E plane, (b) H plane

## 3. CONCLUSIONS

A compact dual-band dielectric resonator antenna fed by a microstrip line has been proposed and measured. The designed prototype has a dual-band operation suitable for both DCS (1710 - 1880MHz) and WLAN (2400 - 2484MHz) applications. The compact dual-band directional antenna achieves a desirable directional radiation pattern with a gain of 4.7 dBi for the 1.8 GHz band and 4.2 dBi for the 2.4 GHz band. It has a small size which satisfies the new communication system requirements.

## ACKNOWLEDGEMENTS

The authors are grateful to several ESEO direction for their financial support during this year research work.

**Authors**

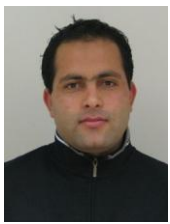

**Hedi RAGGAD** was born in Bengardene, Tunisia. He received his Eng. degree in electronic engineering from Marne La Vallee University France, in 2007.He is currently working toward the Ph.D. degree in ESEO Angers, France, His research interests include dielectric resonator antennas and passive RF devices.

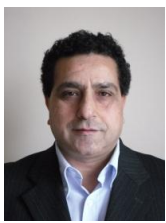

**Mohamed Latrach** was born in Douar ksiba, Sless, Morocco. He received his Ph.D. degree in electronics from the University of Limoges, Limoges, France, in 1990. Presently he is a professor of Microwave Engineering and head of the Radio & Microwave group at the "Ecole Supérieure d'Electronique de l'Ouest, ESEO," Angers, France. His field of research interest is the design of hybrid, monolithic active and passive microwave circuits, metamaterials, Left Handed Materials, antennas, wireless power transmission and their applications in RFID and communication systems, etc.

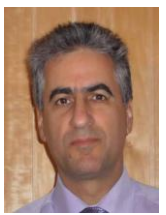

**Tchanguiz Razban** was born in Tabriz, Iran, in 1956. He received the engineering degree from university of Tehran, Iran, in 1980. He received the PHD and Doctorat d'Etat from Institut National Polytechnique de Grenoble (INPG), France, in 1983 and 1989, respectively. He is currently a full professor of electrical engineering at Ecole Polytechnique de Nantes. Since 1999 he has been the chief manager of the electrical engineering department, deputy director and finally director of Ecole Polytechnique until January 2010. His research interests include MMIC's and printed circuits for wireless communications, smart antennas and optical-microwave interfaces.

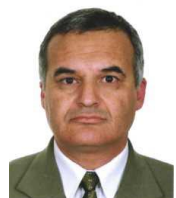

**Gharsallah** received the degree in radio frequency engineering from the Higher School of Telecommunication of Tunis in 1986 and the Ph.D. degree in 1994 from the Engineering School of Tunis. Since 1991, he was with the Department of Physics at the Faculty of Sciences of Tunis. Actually, he is a Full Professor in Electrical Engineering and Director of the Engineering studied in the Higher Ministry Education of Tunisia. His current research interests include smart antennas, array signal processing, multilayered structures and microwave integrated circuits. He has about eighty papers published in scientific journals and more than a hundred papers conferences. Professor Gharsallah supervises more than twenty thesis and fifty Masters.